# Beam Position Monitor Electronics Upgrade for Fermilab Switchyard *

P. Stabile[#], J. S. Diamond, J.A. Fitzgerald, N. Liu, D. K. Morris, P. S. Prieto, J. P. Seraphin

FERMILAB, Batavia, IL, 60510, USA

*Abstract*

The beam position monitor (BPM) system for Fermilab Switchyard (SY) provides the position, intensity and integrated intensity of the 53.10348MHz RF bunched resonant extracted beam from the Main Injector over 4 seconds of spill. The total beam intensity varies from $1 \times 10^{11}$ to $1 \times 10^{13}$ protons.

The spill is measured by stripline beam postion monitors and resonant circuit. The BPMs have an external resonant circuit tuned to 53.10348MHz.

The corresponding voltage signal out of the BPM has been estimated to be between -110dBm and -80dBm.

## INTRODUCTION

During current operation the Main Injector accelerates beam to 120Gev and extracted to the Switchyard (SY). Beam is extracted at the MI52 area, transported through P1, P2 and P3 beam lines, and then steered to the original Switchyard beam line where it traverses Enclosures B, C and the F-manholes, finally arriving to the Target [1].

Extraction is implemented by the half-resonant integer mode and is regulated by a quadrupole circuit (QXR) to resonantly extract beam over 4 seconds.

The typical spill intensities vary from $1 \times 10^{11}$ protons to $1 \times 10^{13}$ protons per machine cycle with the protons distributed into 486 of the 588 53.10348MHz RF buckets that make up the machine's circumference. Extracting $1 \times 10^{13}$ protons over 4 sec accounts for $2.5 \times 10^{12}$ protons per second (pps). With a revolution frequency of the machine equal to 90.312KHz (53.10348MHz/588), a total of $27.68 \times 10^{6}$ protons are extracted per turn ($2.5 \times 10^{12}$ pps/90.312KHz). Dividing the number of protons per turn by the number of buckets a total of $56.95 \times 10^{3}$ protons per bunch can be estimated. Extracting $1 \times 10^{12}$, using the same calculations, accounts for $5.695 \times 10^{3}$ protons per bunch

## SYSTEM DESCRIPTION

The stripline BPMs have a pair of detector plates, facing each other inside the beam pipe.

The plate capacitances are equal to 65pF and the plate-to-plate capacitance is 8pF.

In order to resonate at the desired extraction frequency three inductors were added to the circuit (Figure 1) [2].

During extraction the particle beam can be considered as a current generator at 53.10348MHz (flat top extraction frequency). The signal power detected by the position detector can therefore be increased by raising the detector shunt impedance for the Fourier components of beam current that are chosen (for simplicity the detector is tuned for the n=1 Fourier component).

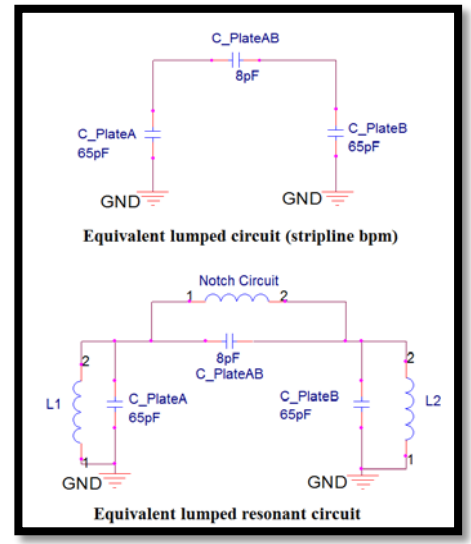

**Figure 1:** Equivalent circuit for resonant BPM.

The detectors are tuned to resonate at the RF frequency of 53.10348 MHz, with a resulting increase of detector shunt impedance from 50Ω to 9.5 KΩ and a Q of around 190.

The original beam position monitor system for the SY was first designed in 1985 [3] and then installed in 1986.

The new designed electronic system consists of:
1. Detectors (Resonant BPMs).
2. RF Transition Board (Analog Front-End).
3. Digitizer (Digital Acquisition and Processing).
4. MVME 5500 Single Board Computer (Software interface).

Figure 2 shows a block diagram of the SY electronic system.

___________________________________________

* This work was supported by the U.S. Department of Energy under contract No. DE-AC02-07CH11359

# stabile@fnal.gov

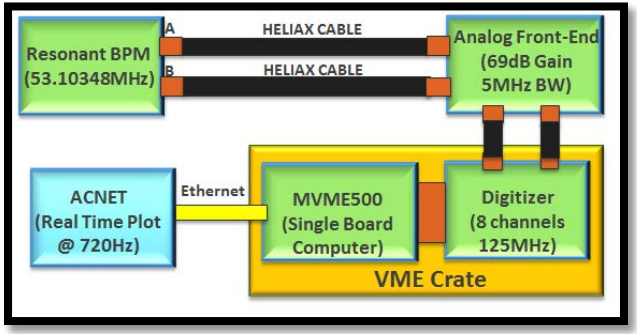

**Figure 2:** SY electronics schematic block.

*Resonant BPMs*

Two different types of resonant BPMs [4] are installed in the beam line:
- A short beam detector (7" plate length, $< \lambda/4$).
- A long detector (1 meter plate length, $\lambda/4$).

Figure 3 and Figure 4 show a view of the short type.

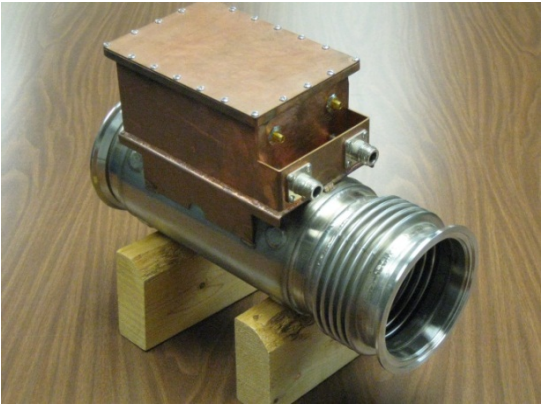

**Figure 3:** Short BPM Detector (1)

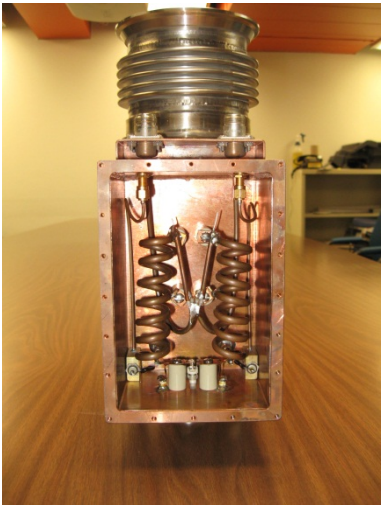

**Figure 4:** Short BPM Detector (2)

The detector body, or beam pipe, is schedule 40 304 stainless pipe of 4.056" I.D. The curved detector plates are copper, supported off the pipe bore by G-10 spacers.

A shorted coax hardline $< \lambda/4$ long is used in the 1 meter detector. In the short detectors, where much more inductance is needed in a limited space, the inductor has been built using a supported copper single-layer solenoid.

Referring to BPM design notes [5] and considering the cylindrical geometry of the detectors, the first harmonic of the output voltage, per electrode, for a centred beam is given by:

$$V_1 = 2.41 \times 10^{-10} N_p \, vrms \quad (1)$$

$N_p$ is the number of protons per bunch. The sensitivity of the BPMs is calculated to be:

$$\left(\frac{V_a}{V_b}\right)_{dB} = 0.67x \quad (2)$$

Where x is the distance the beam is displaced from the detector canter. Actual wire measurements show the sensitivity to be between 0.57dB/mm and 0.63dB/mm.

*Analog Front-End*

The analog front-end is composed by a RF analog board using a cascade of three RF low noise amplifier (total gain of 69dB). Due to the low signal to noise ratio (SNR~12dB) the noise figure of the amplifiers is a key parameter in the design. The amplifier chosen for this application reports by datasheet a noise figure of 0.7dB. A 53.10348MHz band pass filter with a 5MHz bandwidth has been added to the signal path to improve the SNR.

Since the bunch charge range is between $56.95 \times 10^3$ and $5.59 \times 10^3$ protons, the voltage range at the analog front-end input can be estimated to be between 13.72$\mu V_{rms}$ (-84dBm) and 1.38$\mu V_{rms}$ (-106dBm) per plate per bunch. To connect the BPM plates to the analog front-end ½' heliax cables are used.

A prototype of the analog board amplifying the resonant BPM signal is currently being used to measure the intensity and the position of the beam extracted from the Main Injector.

A new transition board is being layed-out. While the current board reads only one BPM, the new one will be able read the signal from 4 BPMs at the same time (8 analog channels). A 6-bit digital attenuator is added to the signal path to compensate differences between channels

The new design will also include a communication interface between the analog board and the crate controller processor using a CAN bus interface.

*Digital Acquisition and Processing*

A digitizer (8 channel, 125MHz, designed at Fermilab) acquires the output signal of the analog front-end.

The signal is first sampled at 122.162935MHz (~2.3 times the fundamental frequency of the beam signal) through a 14 bit ADC and then scaled to 16 bits. The data is then down converted to base band and digitally filtered obtaining the in phase (I) and in quadrature (Q) components (Figure 5). The total pass/stop band of the filter has been designed to be 2KHz/5KHz.

The digital processing architecture is composed by:
- NCO: 32bits, center frequency 53.10348MHz
- Scale Register on 8 bit
- CIC: Decimation by 2048;
- CFIR: 24 Taps, Decimation by 2, Pass band 4KHz, Stop Band 10KHz;
- PFIR: 32 Taps, Decimation by 2, Pass band 2KHz, Stop Band 5KHz;
- 3 Gain blocks;

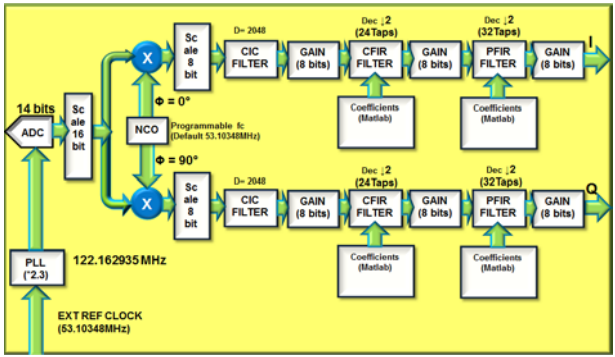

**Figure 5:** Digital Signal Processing Block diagram 1

Figure 6 shows a block diagram of the data processing following the signal processing.

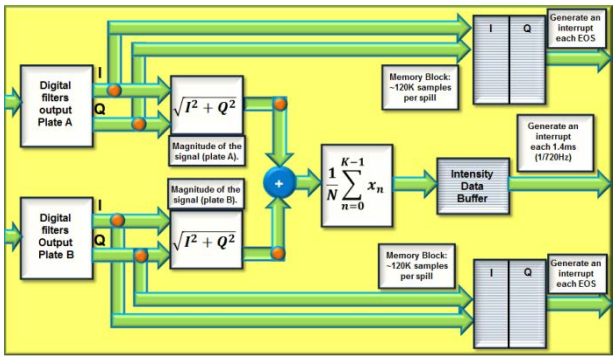

**Figure 6:** Digital Signal Processing Block diagram 2

The throughput of the I and Q generated for each of the two plates of the BPM is 14912K samp/sec. The digitizer board saves I and Q data in a memory buffer and makes them available to be copied and stored in a database by the firmware. The data is automatically overwritten at the next extraction cycle. The calculated square root of sum of squares of I and Q data for each BPM is averaged (N=18) and then saved in a buffer accessible by the main processor. The final update rate to plot is in real time with intensity and position data fixed to 720Hz.

The digital filtering and signal processing is the biggest improvement of the new system over the the old one. A desired signal bandwidth can be re-programmed and modified simply re-loading the FPGA firmware and the intensity and position can be plotted in "real time" during the extraction.

*Software Interface*

One or more digitizers exist in a VME crate with an MVME 5500 Single Board Computer (SBC) running custom software and a Fermilab built clock decoder. To the Fermilab accelerator controls system (ACNET) this collection of hardware and software is known as a "front-end". Several front-ends are required to accommodate a large number of BPMs dispersed over the SY beam line. The front-end computer runs the VxWorks real-time operating system, which is common practice for data acquisition applications at Fermilab. An effort is underway to port the front-end software to Linux.

The primary responsibility of the front-end software is to arm the digitizers before the slow spill begins then read down-converted samples from the digitizers, calculate positions and present them to the control system. Several graphical applications already exist that the user can utilize to display, log and analyze data once it is has been captured by the front-end. Real-time plotting of position and intensity is provided by averaging every 18 down converted samples to match the capability of the control system (720Hz). The full 4 seconds of down converted samples are collected in the digitizer's memory during the slow spill. After the end-of-beam event all of the samples collected from the digitizers in the front end are transferred to the computer's memory and can then be saved to a file or downloaded via the control system for analysis.

## RESULTS

Figure 7 shows the measured beam being extracted (I:702RI) by one the 4 BPMs installed in the P1 line compared to the amount of beam extracted (I:BEAM).

A total efficiency of 93%, I:702EFF (I:702RI/I:BEAM21) for the Main Injector extraction has been measured during spill.

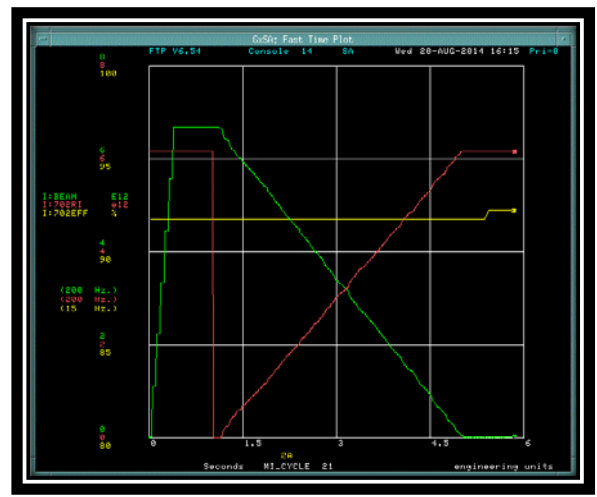

**Figure 7:** Beam extraction intensity measurement.

More tests were run to characterize the resolution of the BPMs. In Figure 8 one of these tests is shown. The total beam intensity in the machine was equal to 5.48e12 and, at the end of the spill, 1.31e11 protons (2.4% of the total beam) were sent to the abort line. A drop of 2% in efficiency was recorded by the intensity monitor.

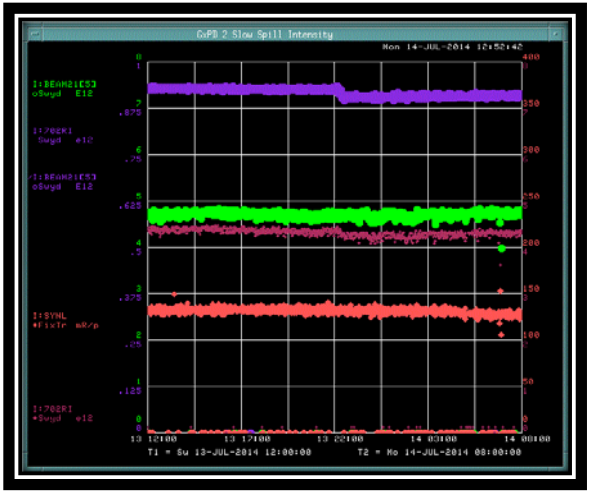

**Figure 8:** Spill efficiency drop of 2%.

The error in resolution is around 0.4 to 0.5% for the intensity measurements.

Figure 9 shows a sweep in the percentage of beam extracted. I:TORO03 is the amount of the beam going to the abort line and I:702RI is the intensity measured.

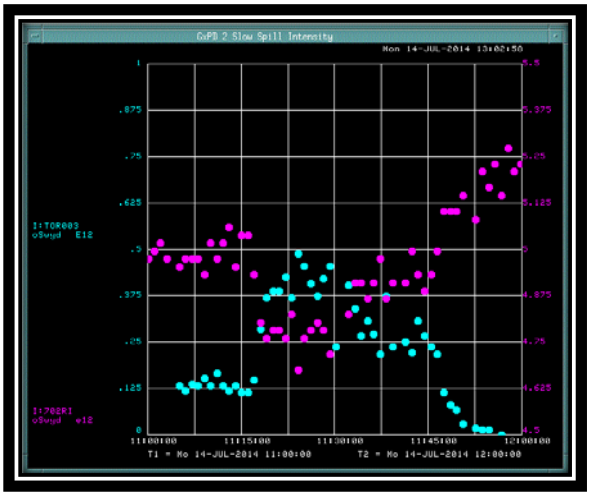

**Figure 9:** Beam extracted vs Beam going to the abort line

To verify the validity of the position recorded by the BPM, a multiwire has been placed in the beam line to compare the results. In Figure 10 the comparison between two vertical BPMs (702 and 714) and the multiwire is shown.

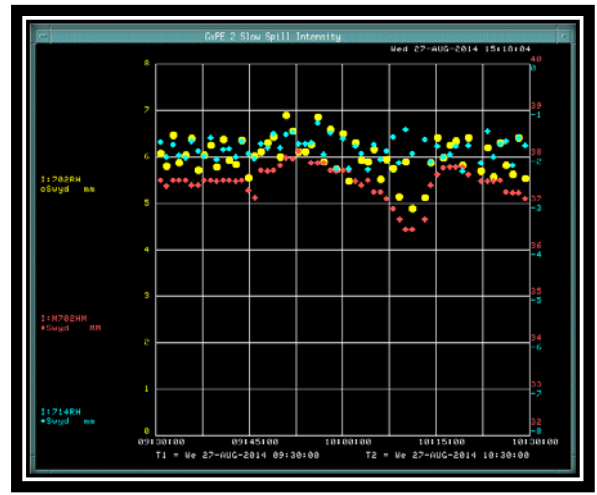

**Figure 10:** Multiwire vs BPMs vertical position comparison.

During these measurements the vertical beam position was deflected. The results show how both vertical BPMs follow the variation in position of the beam and how these results match with the results from the multiwire.

## CONCLUSION

An upgrade of the old SY electronics system has been prototyped and tested. Results show a resolution for the intensity measurement around 0.4/0.5%. The resonant BPM have been proven also good results in position measurement. A new prototype to improve the performances of the system is currently at the layout stage and is expected to be ready for the coming fall.

## ACKNOWLEDGMENT

The authors would like to acknowledge all of the Instrumentation and Main Injector group for their support with this work.

## REFERENCES

[1] Switchyard Rookie Book "Fermilab Doc BD version 3.3".
[2] Q, Kerns et.al., Tuned Beam Position Detectors 1987 Particle Accelerator Conference Session T47, and for his hours on RF theory and techniques.
[3] R. Fuja et. al., Electronics Design for the Fermilab Switchyard beam Position Monitor System Session T21
[4] R. Shafer et. al., Fermilab Energy Doubler Beam Position Detector, IEEE Transactions on Nuclear Science, Vo. NS-28, No. 3, June 1981, pp. 2290-2292.
[5] R. Shafer "BPM Notes" and discussion, guidance and patience.